\documentstyle[iopconf1]{article}
\begin{document}

\title{\bf Cosmological Insights from Supernovae} 

\author{ Pilar Ruiz--Lapuente\dag\ddag\footnote{E-mail:
 pilar@mizar.am.ub.es}}

\affil{\dag\ Department of Astronomy, University of Barcelona, 
Mart\'\i \ i Franqu\'es 1, E--08028 Barcelona , SPAIN}

\affil{\ddag\ Max--Planck Institut f\"ur Astrophysik, 
 Karl--Schwarzschild-Str. 1, D-86460 Garching, GERMANY}

\beginabstract
While low--z Type Ia supernovae
are used to measure the
present rate of expansion of the Universe,
high-z Type Ia measure its variation due to
the cosmic matter--energy content.  Results from those
determinations imply a low matter density Universe with a non--zero 
cosmological constant (vacuum--energy component). The expansion rate
 of the Universe accelerates, according to these determinations.     
The validity of the Type Ia supernova approach for
 this cosmological research is addressed.  An account is
given of additional prospects to further investigate through 
supernovae what the
Universe is made of. Those attempts range from
constraining the large scale dark matter distribution to 
further test and interpret the presence of a vacuum energy component.

\endabstract

\section{Introduction}

 The epoch we are approaching unveils the 
 Universe we live in and sheds new light on some 
 of the problems 
 that have remained open for several decades. 
 The path already followed  
 has been a controversial but exciting  
 enterprise. Only two years ago
 we seemed to be facing a critical moment 
 when comparing the present value of the expansion
 of the Universe, the age of the Universe as derived 
 from the oldest objects, and the preliminary results 
 on the matter content of the Universe. 
 New, better results draw today a different picture,
 which appears reinforced by the agreement in the 
 results from two independent groups in search  of 
 the cosmological values  $\Omega_{M}$ (matter density
 parameter) and $\Omega_{\Lambda}$ (vacuum energy density content
  $\Omega_{\Lambda}$ = $\Lambda$/ 3(H$_{0}$)$^{2}$).
 The results  will be reviewed in detail in the 
 next two contributions (Hook et al. 1998; Leibundgut 1998).

  To test the model of our Universe and derive its expansion rate
  it is necessary
  to choose an object which can be used to trace distances with
  high accuracy. Cepheids
  and their period--luminosity relationship (Leavitt 1912)
  were first used by Hubble and are still used today
  to provide accurate distance determinations.   
  To go to further depths in the Universe and attempt a 
  determination of $\Omega$ and 
  $\Lambda$, we need to have an object whose properties are
  universal (i.e. independent of the time and the region of 
  space), homogeneous (i.e. being a low dispersion sample 
  in brightness, or drawing a sharp relationship with brightness) 
  and observable up to large redshift.
  The thermonuclear explosion of WDs, phenomenologically 
  known as supernovae of Type Ia, turns to be the best
  available astrophysical object in this cosmological context. 
  Studies of those supernovae in our close vicinity,
  at intermediate redshift, 
  and at high z  show that they all have the same 
  properties and fulfill the same set of relationships.

  In the next section, we will describe the empirical 
  relationship, and in the following sections we will 
  elaborate on the underlying method.

\section{Luminosity of Type Ia supernova and H$_{0}$}

\subsection{Brightness--decline}

  The first uses of SNe Ia to determine the 
  cosmological parameters assumed those objects to be
  standard candles, implying that they all had similar 
  luminosities. In 1977 Pskovskii realized a 
  correlation between the brightness at maximum and 
  the rate of decline of the light curve. 
  First explanations attributed such correlation to external
  factors such as interstellar reddening (Boisseau \& Wheeler 1991). 
  However, a systematic follow--up of SNe Ia (Maza et al. 1994;  
  Filippenko et al. 1992a,b; Leibundgut et al. 1992; Phillips 1993, 
  Hamuy et al. 1996a,b) confirmed the brightness--decline relation.  
  Phillips (1993), and Hamuy et al. (1996a,b) 
  quantified it by studying supernovae 
  at z $\sim$ 0.1, a z large enough that
  peculiar motions do not introduce dispersion in the magnitude--redshift
  diagram. The intrinsic variation of SNe Ia is written as a  linear
  relationship of the sort:

\begin{equation}
M_{B} = -19.25 + 0.78\ [\Delta m_{15}(B) - 1.1] + 5 \ log (H_{0} / 65) 
\end{equation}  

\noindent
 with dispersion of only $\sigma \sim 0.17$. 
The relationship 
is calibrated in terms of an arbitrary value of $H_{0}$. 
$\Delta m_{15}$, the parameter of the SNe Ia light curve family,
 is the number of magnitudes 
decreased in 15 days after maximum.

\bigskip
\bigskip



\begin{figure}
\input epsf2
\bigskip
\bigskip
\bigskip
\epsfxsize=210pt
\epsfbox{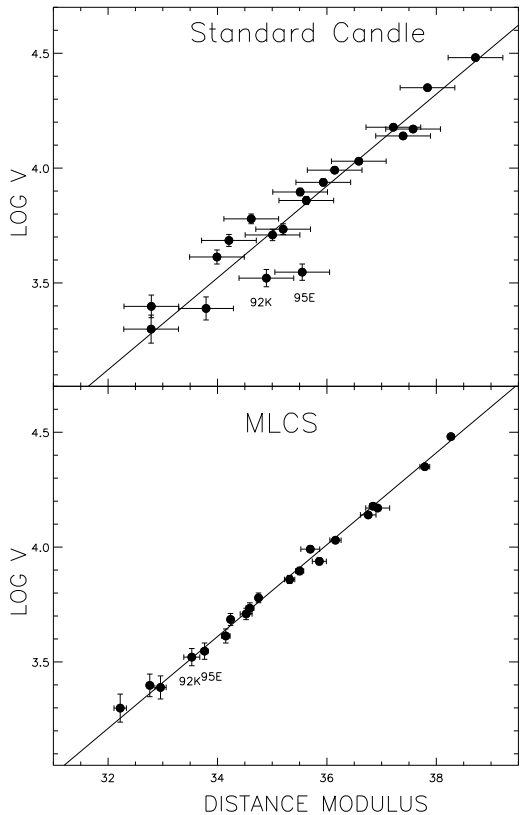}
\vspace{6mm}
\caption{
 The Hubble diagram for SNe Ia before (top panel) and 
 after (bottom panel) correction of the width--absolute
 brightness relationship. The figure is from Riess, 
 Press \& Kirshner (1995a).}
\end{figure}


Other groups have formulated the 
brightness--decline  correlation in a different way. 
 Riess, Press \& Kirshner (1995a,b)
use the full shape of the light curve in respect to a template.
 This is also the formulation used by the High-Z Team 
in their high redshift 
supernova studies. The Supernova Cosmology Project collaboration, 
on the other hand, 
has introduced the {\it stretch--factor, s}
as a parameter to account for the brightness--decline relationship.
 All ways to parameterize the effect 
give equivalent results. The degree at which this correlation reduces 
the scatter in the Hubble diagram is shown in Fig.1.

\bigskip

The relationships above, if used in combination with local 
calibrators, provide a direct measurement of H$_{0}$. 
Those local calibrators are Type Ia supernovae for which 
Cepheid distances to their host galaxies have been determined
through HST observations.
Using the first local available calibrators with distance provided by 
 Cepheids 
 (SN 1937C, SN 1972E, SN 1981B, SN 1990N) and the brightness--decline
 relationship, Hamuy et al. (1996a,b)
derived a H$_{0}$ value of 63.1 $\pm$ 3.4 (internal) $\pm$ 2.9 (external) 
 km s$^{-1}$ Mpc$^{-1}$. 
Riess, Press \& Kirshner (1996) derived a value of 64 $\pm$ 6 
km s$^{-1}$ Mpc$^{-1}$. Whereas most of the empirical determinations of
 H$_{0}$ using SNe Ia give values around 65 (see also the theoretical 
 predictions in the next subsections), 
 Sandage and Tammann (1997), and Saha et al (1997) argue for 
 a H$_{0}$ value $ < $ 65. Sandage, Tammann  
 and collaborators, and 
 Branch and collaborators have selected from the SNe Ia sample
 those presenting bluer colors (``normal SNe Ia'') instead of using 
  the whole sample correlation as in (1). The H$_{0}$ 
 value found is  57 $\pm$ 6 km s$^{-1}$ Mpc$^{-1}$ (for a review,
  see Branch 1998).

  The view held by most authors is that differences on these
  discrepant H$_{0}$ values 
 arise when instead of using the whole sample of SNe Ia 
 and the brightness--decline
 correlation, a restricted use of the subsample of 
``normal'' looking SNe Ia is made. Differences also originate in the
  way to place SNe Ia in the Hubble flow. 
   Further investigation must clarify the still remaining discrepancies.

\bigskip

\subsection{Physical approach: 
Boundaries on $H_{0}$ from  WD explosions}

\smallskip

 The previous paragraphs gave a brief account of the
  empirical calibrations
 of the luminosity of Type Ia supernovae. From a basic understanding 
 of Type Ia supernovae as physical objects, one can also put constraints
 on the value of the Hubble constant. 

\noindent
 The absolute magnitude of a exploded WD 
 has a limit established by the maximum mass of any WD: 
 the Chandrasekhar mass $M_{Ch} \approx 
1.38 M_{\odot} \ ( Y_{e}/ 0.5)^{2}$, where $Y_{e}$ is the number of
 electrons per nucleon.
 A WD accreting mass beyond the Chandrasekhar mass would
  either undergo a gravitational collapse forming a neutron star, or
  it would form a exploding Chandrasekhar mass object plus an envelope 
  of some mass around. The former would be an underluminous explosion,
  and the second would not imply significantly higher
  absolute magnitudes than the bare exploded C+O WD. 

\bigskip 

\noindent
   In the thermonuclear explosion of a Chandrasekhar WD,
  the generated kinetic energy,
  $E_{kin}$ and the radioactive energy $E_{^{56}Ni}$ are linked.
  The larger the radioactive energy from $^{56}Ni$, providing the
  luminosity, the highest expansion velocities $E_{kin}$ have the ejecta.  
  A fast expanding ejecta traps less efficiently the radioactive
  energy: thus a self--constraining interplay on the final absolute 
  magnitude is  
  obtained for explosions of different energies to give a maximum absolute
  magnitude which a Type Ia explosion could achieve. 
  That limit corresponds to a final minimum $H_{0}$ of 
 about 50 km s$^{-1}$ Mpc$^{-1}$.

\bigskip

\noindent
   An upper limit to the value of $H_{0}$ 
  is provided by the discussion of the minimum mass of a exploding WD, and
  if that could correspond to what we see as SNe Ia. The exploration of
  the range of possible exploding WDs by various proposed mechanisms    
  suggests that the range of what we can observe if a whole range of WDs  
  below the Chandrasekhar mass explode is much wider than the objects we
  actually identify as SNe Ia. This argument disfavors $H_{0}$ larger than
  75 km s$^{-1}$ Mpc$^{-1}$.

\subsection{Model light curves and spectra}

\bigskip

  Going a step further in the theoretical examination of 
  Type Ia supernovae, a comparison of calculated light curves
  and observations can led to estimate the absolute luminosity of 
  those explosions and derive H$_{0}$.
   The exploration of light curves  of exploded WDs by  
  Arnett, Wheeler \& Branch (1985), lead to 
  H$_{0}$ $\sim$ 60 km s$^{-1}$ Mpc$^{-1}$. 
 Using the light curves 
 in different colors  and a larger set of explosion models,  
 values of H$_{0}$=67 $\pm$ 7 km s$^{-1}$ Mpc$^{-1}$ are found by
 H\"oflich et al. (1997). 
  Those authors can reproduce the above correlations
  by changing the explosion mode and energy of centrally ignited 
  WD at the Chandrasekhar mass. Eastman \& Pinto (1993), from their
  own light curve calculations, find that 
  the magnitude decline correlation can be explained 
  in different ways, even as a result of different total masses, 
  by exploding sub--Chandrasekhar WDs.

\begin{table}
\begin{center}
\footnotesize\rm
\caption{A table showing the mean absolute brightness for Type Ia
 SNe, and H$_{0}$.}
\begin{tabular}{llll}
\topline
 Source  & H$_{0}$ km s$^{-1}$ Mpc$^{-1}$ & $<M_{B}>$  & Method \\
\midline
 Saha et al. 1997 & 58$_{-8}^{+7}$   & -19.52 $\pm$ 0.07 & SNeIa/Cepheids \\
  Hamuy et al. 1997 & 63 $\pm$ 6 & -19.28 $\pm$ 0.1 & 
SNe Ia/$\Delta m_{15}$/Cepheids \\
  H\"oflich \& Khokhlov 1997 & 67 $\pm$ 7 & -19.45 $\pm$ 0.2
 & SNe Ia Light Curves \\
  Ruiz-Lapuente 1996 & 68 $\pm$ 13 & -19.3 $\pm$ 0.3 & Late SNe Ia 
spectra \\          

\bottomline
\end{tabular}
\end{center}
\end{table}

\noindent
  Through the prediction of spectra of Type Ia supernovae,  
  which are good discriminators of density and temperature,
  and comparison with observations, 
 H$_{0}$ is found to be 68 $\pm$ 6 (stat) $\pm$ 7 (sys)
  km s$^{-1}$ Mpc$^{-1}$   (Ruiz--Lapuente 1996). 

\bigskip

\begin{table}
\begin{center}
\footnotesize\rm
\caption{A comparison of distance measurements for two SNe Ia
 by three methods used to determine H$_{0}$ with SNe Ia.}
\begin{tabular}{llll}
\topline
 Supernova/galaxy & Theor. light curves & Theor. Late Spectra & 
 Cepheids \\

                  &   m-M  &     m-M         &  m-M  \\
\midline
 SN 1990N / NGC 4539  & 31.5 $\pm$ 0.4 $^{a}$ &  31.81
 $\pm$ 0.2  $^{b}$ &  32. $\pm$ 0.23 $^{c}$ \\
 SN 1989B / NGC 4532  &  29.7$^{+0.64}_{-1.07}$ $^{a}$ & 
  29.95 $\pm$ 0.4 $^{b}$  & 30.37 $\pm$ 0.16 $^{d}$ \\
\bottomline
\end{tabular}
\centerline{}
\centerline{
$^{a}$  H\"oflich et al. 1996; $^{b}$  Ruiz--Lapuente 1996; 
$^{c}$  Sandage et al. 1997; $^{d}$  Tanvir et al. 1995}
\end{center}
\end{table}

\noindent
   When comparing the luminosities of individual SNe Ia 
  obtained by different methods (theoretical light curves, spectra,
  or through the empirical method) the results agree 
  up to a high extent (see Table 1). This illustrates the convergence 
  achieved during the last years. Empirical and theoretical
  results on H$_{0}$ from SNe Ia,
  give values centered at 
  65 km s$^{-1}$ Mpc$^{-1}$, with up to 
  10$\%$ disagreement between various determinations (Table 1 and 2).
  Table 2 shows that the methods applied to determine distances to
  SNe Ia give consistent results.

\section{High-z Type Ia supernovae}

\bigskip
 
   The previous section dealt with the absolute calibration
  of the luminosity of SNe Ia and the determination of the 
  present value of the expansion of the Universe (H$_{0}$).   
 Independently of that calibration and independently of the
 H$_{0}$ derived from it, one can use the apparent
 brightness of Type Ia
 at high z (z $\ >$ 0.3) to measure the global curvature
 of the Universe. This has been the scope of the Supernova
 Cosmology Project (see Hook et al. in this volume) 
 and of the High--z Team (Leibundgut in this volume). The approach
 relies in that SNe Ia have observational 
 properties which are well--understood
 and identical at all redshifts.

\begin{figure}[hbtp]
\input epsf
\centerline{\epsfysize12cm \epsfbox{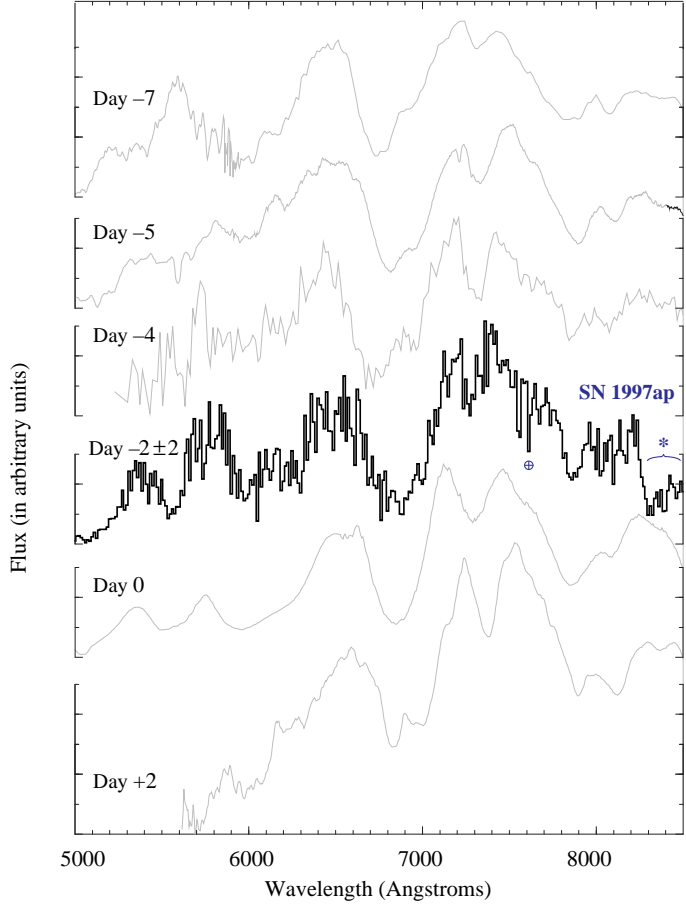}}
\nopagebreak[4]
\vspace{4mm}
\caption{
 Spectrum of SN 1997ap at z=0.83 is shown in the 
 rest frame compared with the spectra of nearby SNe Ia.
 The spectrum allow a classification of the supernova within the 
 family of SNe Ia (Perlmutter et al. 1998;
 Nugent et al. 1996).}
\label{redu}
\end{figure}

\smallskip

   Indeed, the exploration of spectra and light curves 
   of SNe Ia shows that the high--redshift supernovae have 
   light curve shapes and spectra like the low--redshift ones
   (Goldhaber et al. 1998; Perlmutter et al. 1998). It is  
   possible to determine for each high-z SN Ia where does it lie
   within the family of SNe Ia studied at low redshift (see Figure 2).
   This is 
   possible both spectroscopicaly and through their light curves.
   Lightcurve  declines in terms of the 
   stretch factor are measured in high-z SNe Ia 
   and the correlation between brightness 
   and rate of decline is the same as at low z. On the other hand, 
   the family of SNe Ia form a sequence of highly resembling spectra 
   with subtle changes in some spectral features correlated with the 
   light curves shapes (Nugent et al. 1998). 

\begin{figure}[hbtp]
\input epsf
\centerline{\epsfysize8cm \epsfbox{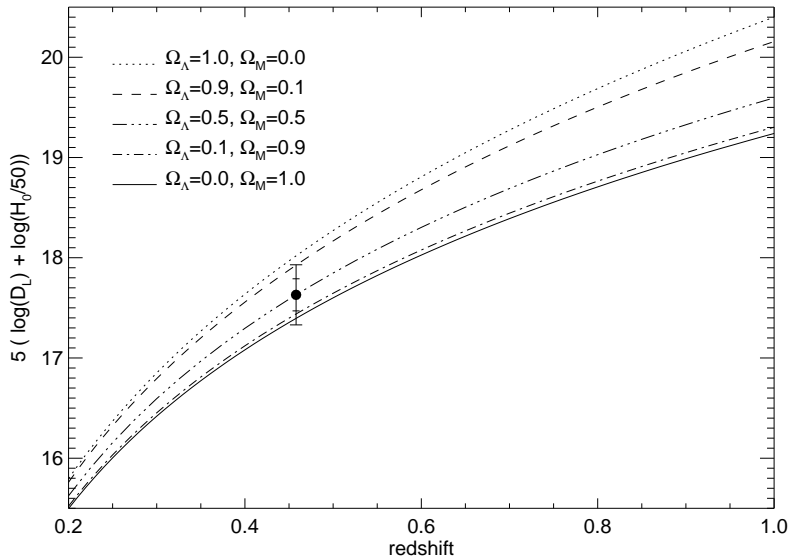}}
\nopagebreak[4]
\vspace{4mm}
\caption{The m(z) relation for 
 different choices of $\Omega_{M}$ and $\Omega_{\Lambda}$,
 compared with an observed SN Ia with z $\simeq$ 0.45, 
 according to Goobar \& Perlmutter (1995). 
 SNe Ia at different z help to discriminate between the
 possible Universe models. 
}
\label{redu}
\end{figure}

\smallskip

   In the following we will address the determination of 
  $\Omega_{M}$ and $\Omega_{\Lambda}$ with high--z SNe Ia.

\bigskip

\subsection{
 $\Omega_{M}$, $\Omega_{\Lambda}$ and the equation of state of the Universe }

   The expansion rate of the Universe  would 
  slow down or speed up depending on its  matter--energy content,
 as formulated in the Einstein field equations (Einstein 1917). It would 
  accelerate if the vacuum energy, i.e 
  cosmological constant term,  would dominate. 
  The first derivations of the luminosity distance of standard candles 
  as a function of z used to combine the matter density term 
   $\Omega_{M}$ (density parameter) and the  $\Omega_{\Lambda}$  term
   (vacuum energy density term,  
  $\Omega_{\Lambda}$ = $\Lambda$/ 3(H$_{0}$)$^{2}$)
   in 
  the so--called deceleration parameter (q$_{0}$=1/2 $\Omega_{M}$-
  $\Omega_{\Lambda}$). The value of  q$_{0}$ would determine 
  whether the expansion of the Universe accelerates or decelerates. 
     Goobar \& Perlmutter (1995) showed that 
  it is better to preserve the terms involving $\Omega_{\Lambda}$
  and $\Omega_{M}$  separately in the luminosity distance expression
  for a standard candle  as a function of z, since they enter with a 
   different z dependence.

   By using the magnitude--redshift relation m(z) as
   a function of $\Omega_{M}$ and  $\Omega_{\Lambda}$
  with a sample
  of high-z SNe Ia of different z, it is possible to determine  
  the region in the $\Omega_{M}$--$\Omega_{\Lambda}$ plane
   favored by the observations (Goobar \& Perlmutter 1995). 
 The relation goes as:

 \begin{equation}
 m(z) = M + 5\ log\ d_{L} (z, \Omega_{M}, \Omega_{\Lambda}) - 5\ log\ H_{0} +
 K_{c} + 25
 \end{equation}

\noindent
 where M is the absolute magnitude of 
 the supernova, $d_{L}$ the luminosity distance, and
 $K_{c}$ is the K--correction (see, for instance, Goobar \& Perlmutter 
 1995). The method is illustrated in Figure 3.

\noindent
   The Supernova Cosmology Project and the High--z team have presented 
   the results of this analysis. For first time, it is found evidence for
   a non--zero $\Lambda$  term.

\noindent
   Steinhardt (1996)  has pointed out that the
   acceleration of the expansion as due to a $\Lambda$  term 
   could  also be explained by  a
   component with a negative pressure, 
   whose 
   equation of state  $p=w \rho$ could have any w within   
   $-1 \ <  w \ < 0$. Such a component, $\it quintessence$ (or Q-component), 
   would be the energy associated with a slowly evolving--down 
   scalar field, which would be responsible for the observed acceleration
   in the rate of expansion of the Universe.
   The vacuum energy $\Lambda$ case is the simplest case with w = --1. 
   Steinhardt (1996) shows how  the m(z)  solutions with 
   non--zero cosmological constant 
   $\Lambda$ and low $\Omega_{M}$ could be mistaken for 
   some solutions where $\Lambda$ vanishes and there is a 
   significant quintessence contribution.
   Huterer \& Turner (1998) propose to 
   use luminosity distance to high--z SNe Ia to put 
   constraints on the quintessence potential. 
     
\bigskip

\subsection{SNe Ia as cosmological candle: evolution?}

\bigskip

  Before bringing SNe Ia to enlighten early Universe physics,
  one may think about the present
  understanding of the systematic effects
  in those cosmological projects.  

\smallskip

\noindent
 Are SNe Ia the universal candles we think they are? 

\smallskip

\noindent
  So far, the Supernova Cosmology Project has found that the
  most distant SNe Ia in their sample are indistinguishible 
  from the nearby ones (see Figure 2).  The use of those SNe Ia
  gives robustness to the $\Omega$ and $\Lambda$ results.
 
\noindent
  Besides the empirical assessment expected to grow in the next years,
  it is worth to debate 
  the evolution in z of SNe Ia properties according to theoretical
  expectations.

\smallskip
   
\subsubsection{Exploration of metallicity effects}

    H\"oflich, Wheeler \& Thielemann (1998) have 
   explored the possible outcome that various initial metallicities and
   ages of the WD could have in the final explosion. 
   This investigation finds that changing the initial metallicity Z
   from Population I to II alters the isotopic composition of the
   outer layers of the supernova ejecta. 
   The influence in the visual and blue light curves is found to be
   small (0.01--0.02 mag). A second effect of the age of the WD comes 
   in the variation of C/O ratio of the WD. A decreased C/O ratio reduces 
   the explosion energy and $^{56}$Ni production. The effect 
  in the light curve is similar to a 
   reduction in the deflagration to detonation transition density.

\noindent  
   As long as the brightness--decline relationship (1) 
   does not change with increasing z, the  cosmological
   uses of SNe Ia are on safe ground.  
   Somehow, it could be that all observable effects reduce to 
   the well--known brightness--decline relationship.

\subsubsection{Exploration of age--mass evolutionary effects}

    Within the present discussion on the evolutionary path that leads
    to Type Ia supernovae, Ruiz--Lapuente, Canal \& Burkert (1997)
    have predicted a number of behaviors that would be expected 
    when going to high redshift, depending on the way the WD is brought to
    explosion (sub--Chandrasekhar or Chandrasekhar explosions).
    A population--age effect is expected because more 
    massive SNe Ia--progenitor binary systems will be selected for
    in younger populations (Ruiz--Lapuente, Burkert \& Canal 1995). 
    If the path to explosion is to reach the Chandrasekhar mass
    through accretion of C--O from a disrupted WD companion 
    (merging WDs scenario), the range of differences in mass of the system
    before starting the accretion that would lead to explosion is 
    of only 5 $\%$. Within this scenario, that mass difference can 
    not lead to a significant effect. If the WD gains mass
    by accreting H from a non-WD companion and burning it to He and C--O, 
    then the typical age of the system when it reaches explosion is
     longer (5 10$^{9}$ yr) and a very small mass variation between 
     z=0 and z=0.8 is found (less than 3$\%$). 
    
\noindent
      All the above suggests that we have one of the best 
     distance indicators one could have. An object whose evolutionary
     effects going back half the age of the Universe are negligible
     because that is the typical time that the star takes to explode. 

\noindent
    At the moment, to complete the sample of SNe Ia at various z 
   is a prioritary goal of the project. That should allow to size
   the systematic uncertanties related to possible evolutionary effects, 
   and to discriminate $\Lambda$ from other contributions
   in the m(z) diagram. 
   
\begin{figure}[hbtp]
\input epsf
\centerline{\epsfysize8cm \epsfbox{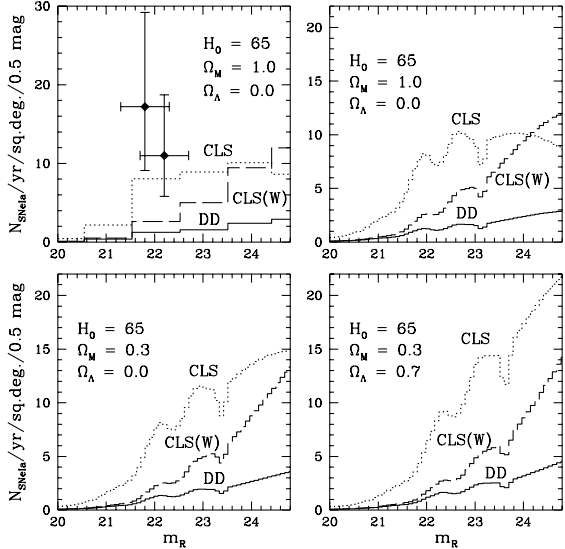}}
\nopagebreak[4]
\vspace{4mm}
\caption{ The effect of a non--zero 
$\Omega_{\Lambda}$ in the number counts
 of SNe Ia is to increase those counts at z close to 1. This effect can 
only be appreciated if the global SFR is known to a precision better 
 than a factor 1.5.}
\label{redu}
\end{figure}

\section{Further cosmological tests with SNe Ia}

  SNe Ia can also be used to study inhomogeneities in the
  matter distribution. Gravitational lensing of  supernovae 
  increase
  the dispersion of the SNe Ia magnitude--redshift diagram
  due to magnification of supernova brightness. With good statistics
  of supernovae it may become possible to measure the magnification
  distribution of the supernovae from observations, and thus determine
  the matter distribution. Both, continuous matter distribution, 
  like medium scale distribtion and large scale distribution, and 
  point--like distribution, like the mass concentration in  
  supermassive compact objects can be examined (Holz \& Wald 1997,
  Metcalf \& Silk 1998, Wambsganss et al. 1997). 
  Constraints on the presence of 
  dark matter halos in the Universe up to high z can 
  be drawn from the SNe Ia results (Metcalf 1998).

\noindent
  Type Ia supernovae at z$\sim$ 0.1  can also be used to trace 
 cosmic flows (Riess, Press \& Kirshner 1995b). By
 probing large scale velocity fields in the Universe an 
 independent  measure of $\Omega^{0.6}/b$ (i.e. a combination of
 the density parameter and the bias factor which informs on how 
 galaxies trace the mass distribution) can be obtained.

\smallskip

\noindent
    An independent test on the presence of $\Lambda$ comes from the
   counts of SNe Ia (number of SNe Ia yr $^{-1}$ sqdeg$^{-2}$
   dz$^{-1}$) as shown in Figure 4. The different comoving volume
   at a given z for Universe models with a non--zero $\Lambda$ and 
   models with $\Lambda$=0 predicts that 
   the number of SNe Ia at z $\sim$ 1 would
   increase more rapidly in $\Lambda$ dominated universes 
   (Ruiz--Lapuente \& Canal 1998). The Supernova Cosmology Project
   is obtaining new values of those numbers by carefully evaluating 
   the results of the searches at various z (Pain et al.1996).

\smallskip

\noindent
   All the precedent is a summary of the cosmological uses of SNe Ia
   undertaken recently, making of supernovae a field which has 
   diversified its cosmological scope. 
   Hopefully, the interesting prospects here mentioned will bring us 
   definitive answers in the near future. 

\smallskip

\noindent
  This work is supported by the Spanish DGES under grant PB94-0827-C02-01. 
  It is a pleasure to thank the organizers of DARK98 for
  providing the opportunity to discuss this and many other interesting
  topics.

\end{document}